\documentclass[sigconf, screen, nonacm]{acmart}
\microtypesetup{expansion=false,protrusion=false}

\AtBeginDocument{%
  }

\renewcommand\footnotetextcopyrightpermission[1]{}

\usepackage{newunicodechar}
\newunicodechar{−}{\textminus}

\usepackage{tabularx,array}
\usepackage{multirow}
\newcolumntype{Y}{>{\centering\arraybackslash}X}

\usepackage{amssymb}

\usepackage{xcolor}
\usepackage{tcolorbox}
\tcbuselibrary{skins}
\usepackage{enumitem}

\usepackage{placeins}
\usepackage{float}

\usepackage{tikz}
\usepackage{pgfplots}
\pgfplotsset{compat=1.18}

\definecolor{questionbg}{RGB}{232, 245, 253}
\definecolor{questionborder}{RGB}{30, 136, 229}
\definecolor{hplusbg}{RGB}{232, 245, 233}
\definecolor{hplusborder}{RGB}{67, 160, 71}
\definecolor{hminusbg}{RGB}{255, 235, 238}
\definecolor{hminusborder}{RGB}{229, 57, 53}
\definecolor{graybg}{RGB}{245, 245, 245}
\definecolor{grayborder}{RGB}{120, 120, 120}

\tcbset{%
  enhanced,
  boxrule=1.2pt,
  arc=2pt,
  fonttitle=\bfseries\small,
  colbacktitle=white,
  coltitle=black,
  left=2mm, right=2mm, top=1mm, bottom=1mm,
}

\title{Ruling Out to Rule In: Contrastive Hypothesis Retrieval for Medical Question Answering}

\author{Byeolhee Kim}
\affiliation{%
  \department{Department of Information Medicine}
  \institution{Asan Medical Center}
  \streetaddress{88, Olympic-ro 43-gil, Songpa-gu}
  \city{Seoul}
  \postcode{05505}
  \country{Republic of Korea}
}

\author{Min-Kyung Kim}
\affiliation{%
  \department{Department of Computer Science}
  \institution{Yonsei University}
  \streetaddress{50, Yonsei-ro, Seodaemun-gu}
  \city{Seoul}
  \postcode{03722}
  \country{Republic of Korea}
}
\affiliation{%
  \institution{INMED DATA}
  \streetaddress{88, Olympic-ro 43-gil, Songpa-gu}
  \city{Seoul}
  \postcode{05505}
  \country{Republic of Korea}
}

\author{Young-Hak Kim}
\affiliation{%
  \department{Division of Cardiology, Department of Information Medicine}
  \institution{Asan Medical Center, University of Ulsan College of Medicine}
  \streetaddress{88, Olympic-ro 43-gil, Songpa-gu}
  \city{Seoul}
  \postcode{05505}
  \country{Republic of Korea}
}

\author{Tae-Joon Jeon}
\authornote{Corresponding author.}
\affiliation{%
  \department{Department of Information Medicine}
  \institution{Asan Medical Center, University of Ulsan College of Medicine}
  \streetaddress{88, Olympic-ro 43-gil, Songpa-gu}
  \city{Seoul}
  \postcode{05505}
  \country{Republic of Korea}
}

\renewcommand{\shortauthors}{Kim et al.}

\begin{abstract}
Retrieval-augmented generation (RAG) grounds large language models in external medical knowledge, yet standard retrievers frequently surface hard negatives that are semantically close to the query but describe clinically distinct conditions.
While existing query-expansion methods improve query representation to mitigate ambiguity, they typically focus on enriching target-relevant semantics without an explicit mechanism to selectively suppress specific, clinically plausible hard negatives.
This leaves the system prone to retrieving plausible mimics that overshadow the actual diagnosis, particularly when such mimics are dominant within the corpus.
We propose Contrastive Hypothesis Retrieval (CHR), a framework inspired by the process of clinical differential diagnosis. CHR generates a target hypothesis $H^+$ for the likely correct answer and a mimic hypothesis $H^-$ for the most plausible incorrect alternative, then scores documents by promoting $H^+$-aligned evidence while penalizing $H^-$-aligned content.
To prevent the penalty from erasing target-relevant evidence when the two hypotheses semantically overlap, we further introduce an \emph{adaptive} contrastive weight that modulates the penalty as a function of the cosine similarity between $H^+$ and $H^-$.
Across three medical QA benchmarks, four retrievers spanning dense (MedCPT, BGE-large-en-v1.5, E5-large-v2) and sparse (BM25) families, and three answer generators, CHR outperforms all five query-expansion baselines and generalizes across retriever families.
On the $n=587$ pooled cases where CHR answers correctly while embedded hypothetical-document query expansion does not, 85.2\% have no shared documents between the top-5 retrieval lists of the two methods, consistent with substantive retrieval redirection rather than light re-ranking of the same candidates.
Adaptive weighting further improves accuracy by up to 4.8 percentage points on average over the fixed weight, confirming that dynamically modulating the mimic penalty resolves the semantic co-occurrence collapse we previously identified as a failure mode.
By explicitly modeling what to avoid alongside what to find, CHR bridges clinical reasoning with retrieval mechanism design and offers a practical path to reducing hard-negative contamination in medical RAG systems.
\end{abstract}



\begin{document}

\maketitle

\section{Introduction}

Retrieval-augmented generation (RAG) has become the dominant paradigm for grounding large language models (LLMs) in external knowledge, particularly in high-stakes medical domains where factual accuracy is critical \citep{lewis2020retrieval, guu2020realm, fan2024survey}.
However, standard RAG pipelines are systematically vulnerable to hard negatives---documents that share surface-level similarity with the query but describe clinically distinct conditions---in medical question answering.
Previous work has documented retrieval relevance as low as 22\% on medical benchmarks \citep{xiong2024benchmarking}, indicating that the main bottleneck lies in retrieval rather than generation.

Query-expansion methods such as HyDE \citep{gao2023precise}, Query2Doc \citep{wang2023query2doc}, CSQE \citep{lei2024csqe}, and ThinkQE \citep{lei2025thinkqe} bridge the query-document semantic gap by generating hypothetical or augmented content.
While these methods enrich the query representation, they share a fundamental limitation.
They model only \emph{what to find} without any mechanism to explicitly suppress \emph{what to avoid}.
When the generated content converges toward a plausible but incorrect interpretation, which is common in medical differential diagnosis, the expanded query remains biased toward the wrong semantic region.

We propose Contrastive Hypothesis Retrieval (CHR), a framework inspired by clinical differential diagnosis that operationalizes this principle at inference time.
CHR produces a contrastive pair consisting of a target hypothesis ($H^+$) describing the likely correct answer and a mimic hypothesis ($H^-$) describing the most plausible incorrect alternative.
The retriever scores documents by promoting $H^+$-aligned content while penalizing $H^-$-aligned content, steering the query away from the hard-negative region and surfacing documents that conventional methods fail to reach.

A naive fixed penalty is not always safe: when the mimic hypothesis semantically overlaps the target hypothesis, subtracting its embedding erases target-relevant signals as well.
To address this, we introduce an \emph{adaptive} contrastive weight that dampens the penalty as the cosine similarity between $\vec{H^+}$ and $\vec{H^-}$ increases, so that highly separable hypotheses receive full contrastive filtering while near-collapse pairs revert to $H^+$-driven retrieval.
Empirically, adaptive weighting recovers accuracy on the failure cases we previously attributed to semantic co-occurrence collapse without sacrificing performance elsewhere.

We evaluate CHR under a broader empirical protocol than prior work.
Beyond three medical QA benchmarks and three answer generators, we test CHR across four retrievers spanning both dense (MedCPT, BGE-large-en-v1.5, E5-large-v2) and sparse (BM25) families, and against five query-expansion baselines.
Across all configurations, CHR outperforms every baseline in aggregate, and retrieval shift analysis on the $n=587$ pooled cases where CHR answers correctly but HyDE does not shows that 85.2\% have zero top-5 overlap, confirming that CHR redirects retrieval to different evidence rather than incrementally re-ranking the same candidate pool.

Our contributions are as follows.
\begin{itemize}[leftmargin=*, itemsep=2pt, topsep=2pt]
    \item We identify a systematic failure mode shared by existing query-expansion methods for medical QA. They model only what to find, leaving the retriever vulnerable to hard negatives when the dominant corpus pattern favors a clinically plausible mimic.

    \item We propose CHR, a contrastive retrieval method that generates a hypothesis pair ($H^+$ and $H^-$) and retrieves documents through explicit contrast, mirroring clinical differential diagnosis. We also introduce an adaptive contrastive weight that softens the mimic penalty when $H^+$ and $H^-$ semantically overlap, mitigating semantic co-occurrence collapse.

    \item We demonstrate consistent improvements over five baselines across three medical QA benchmarks and three generators, and establish that CHR generalizes across four retrievers spanning dense and sparse families. Retrieval shift analysis confirms that CHR redirects retrieval toward fundamentally different semantic regions rather than re-ranking a shared candidate pool.
\end{itemize}

\section{Related Work}
\subsection{Retrieval-Augmented Generation}
Retrieval-augmented generation (RAG) conditions language model outputs on documents retrieved from an external corpus, reducing hallucination and enabling access to specialized knowledge beyond the model's parametric memory \citep{lewis2020retrieval, guu2020realm}.
In medicine, RAG has been widely adopted to ground clinical question answering in biomedical literature \citep{fan2024survey}.
However, \citet{xiong2024benchmarking} show that retrieval relevance on medical benchmarks can be as low as 22\%, indicating that the primary bottleneck lies not in generation but in retrieval quality.
Domain-specific retrievers such as MedCPT \citep{jin2023medcpt}, pretrained on PubMed search logs, improve biomedical retrieval but do not explicitly address the hard-negative problem where clinically distinct conditions share similar surface features.
Alongside domain-specific retrievers, strong general-purpose dense encoders such as BGE \citep{xiao2024bge} and E5 \citep{wang2022text} and classical sparse retrievers such as BM25 remain competitive baselines in medical settings, and a robust retrieval framework should generalize across all three families rather than being tuned to a single retriever.
A complementary line of work focuses on post-retrieval refinement or iterative feedback loops to improve document quality.
These approaches range from rationale-based filtering of medical snippets \citep{sohn2025rationale} and iterative self-reflection for clinical QA \citep{ryan2026selfmedrag} to critique-gated document selection in adaptive RAG systems \citep{asai2024selfrag}.
While such techniques enhance downstream grounding, they operate on a pre-defined candidate pool and cannot rectify fundamentally misdirected searches.
In contrast, CHR addresses this bottleneck at the query construction stage, reshaping the initial retrieval vector to preemptively steer away from anticipated hard negatives.

\subsection{Query Expansion for Retrieval}
Query expansion methods bridge the semantic gap between short queries and long documents by augmenting the original query with LLM-generated content.
HyDE \citep{gao2023precise} generates multiple hypothetical answer documents and averages their embeddings as the retrieval query, relying on noise dilution to filter spurious directions.
Query2Doc \citep{wang2023query2doc} appends a generated pseudo-document directly to the query for sparse or dense retrieval.
CSQE \citep{lei2024csqe} combines corpus-originated pivotal sentences with LLM-generated expansions via a two-stage pipeline, while ThinkQE \citep{lei2025thinkqe} introduces an iterative thinking-based process to refine queries using retrieval feedback.

Although these methods effectively enrich query semantics, they share a fundamental limitation.
They lack a dedicated mechanism to counteract misleading retrieval directions.
This absence of negative constraints leaves the retriever vulnerable to hard negatives, particularly when an expanded query converges toward a plausible but incorrect interpretation---a frequent challenge in medical differential diagnosis.
This necessitates a retrieval framework that can explicitly counteract these biases at the query construction stage.

\section{Methodology}

\begin{figure}[t]
  \centering
  \includegraphics[width=\linewidth]{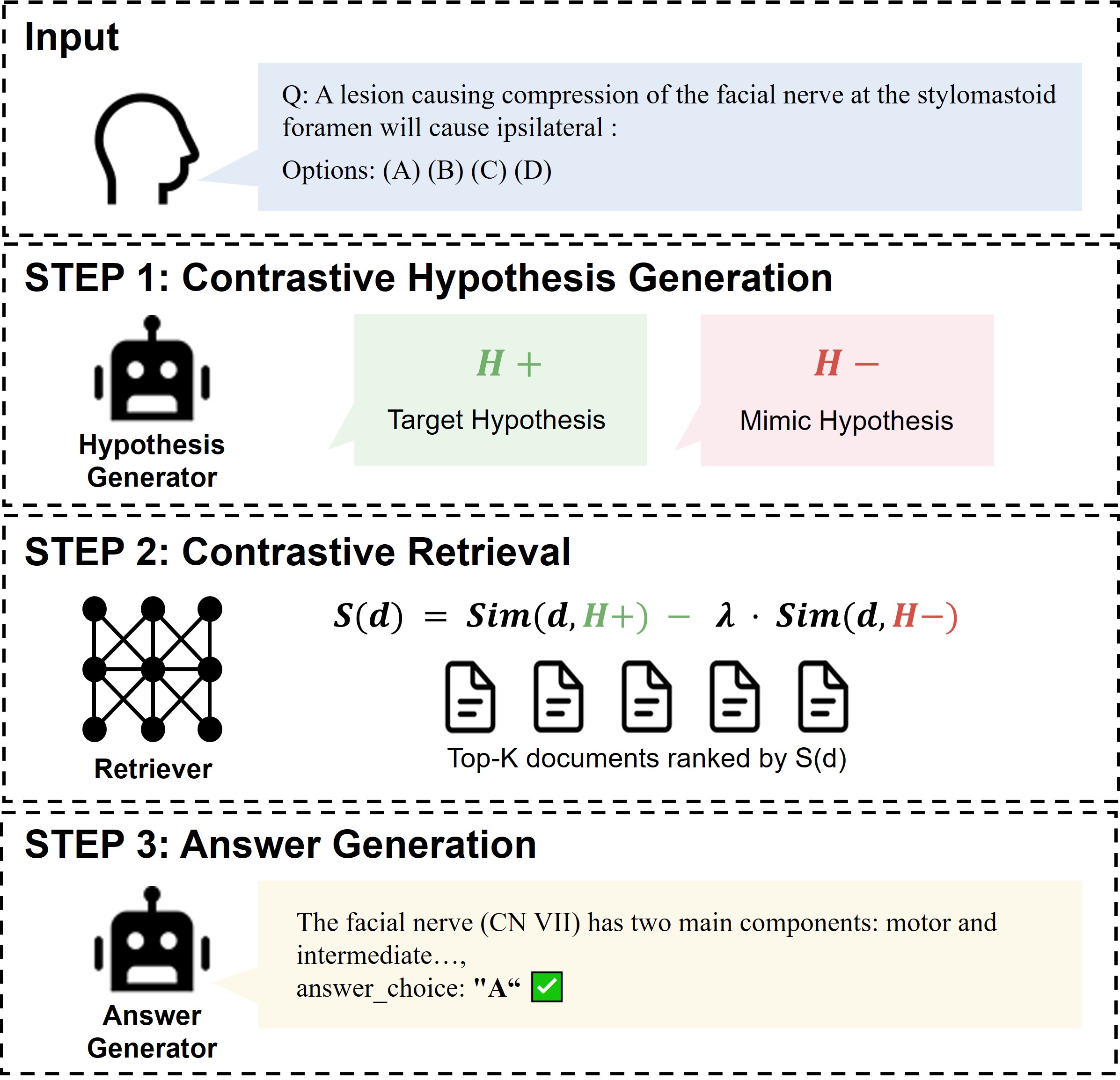}
  \caption{Overview of Contrastive Hypothesis Retrieval (CHR). Given a clinical question, CHR generates a contrastive hypothesis pair consisting of a target hypothesis ($H^+$) and a mimic hypothesis ($H^-$), retrieves documents using contrastive scoring that promotes target-aligned content while penalizing mimic-aligned content, and generates the final answer from the retrieved evidence.}
  \label{fig:pipeline}
\end{figure}

Given a clinical query $Q$ and candidate options $O$\linebreak$= \{o_1, o_2, \ldots, o_n\}$, the objective is to identify the correct answer.
As illustrated in Figure~\ref{fig:pipeline}, CHR operates in three stages.
\textbf{(1)~Contrastive Hypothesis Generation.} Given $Q$ and $O$, an LLM produces a target hypothesis $H^+$ describing the likely correct answer and a mimic hypothesis $H^-$ describing the most plausible incorrect alternative, both in a single structured call.
\textbf{(2)~Contrastive Document Scoring.} Each candidate document is scored by $S(d) = \text{Sim}(d, H^+) - \lambda \cdot \text{Sim}(d, H^-)$, promoting target-aligned evidence while suppressing mimic-aligned hard negatives.
The top-$K$ documents are returned.
\textbf{(3)~Answer Generation.} The retrieved documents are provided to an answer generator along with the original question to produce the final answer.

\subsection{Contrastive Hypothesis Generation}
The core idea of CHR is that suppressing hard negatives requires explicit knowledge of what makes them misleading.
We instantiate this principle through a contrastive hypothesis pair that captures both the target and its closest mimic.

The target hypothesis $H^+$ describes the clinical reasoning and distinguishing features of the likely correct diagnosis.
It captures the pathophysiology, characteristic symptoms, and gold-standard treatment that differentiate the target condition from alternatives.
This hypothesis guides the retriever toward documents containing target-relevant evidence.

The mimic hypothesis $H^-$ describes the most plausible incorrect alternative that shares surface-level similarities with the correct answer.
It articulates why a clinician might mistakenly consider this condition while specifying the subtle features that rule it out.
By explicitly modeling the mimic, CHR enables the retriever to identify and down-weight documents that would otherwise rank highly due to superficial similarity.
This mirrors clinical differential diagnosis, where physicians consider and exclude similar conditions before confirming a diagnosis.

Both hypotheses are generated in a single LLM call using a structured prompt that requests a JSON-formatted output containing $H^+$ and $H^-$.
The prompt instructs the model to analyze the clinical scenario and generate two conflicting hypotheses for retrieval, as shown in Figure~\ref{fig:prompt_template}.

\begin{figure}[t]
\centering
\begin{tcolorbox}[
  colback=gray!5,
  colframe=gray!50,
  boxrule=0.5pt,
  title={\textbf{System Prompt}},
  fonttitle=\small,
  coltitle=black,
  colbacktitle=gray!20
]
\small
You are a medical specialist assisting with complex clinical decision-making. Your goal is to generate precise diagnostic hypotheses to guide an evidence-based search engine. Ensure all outputs are in strict JSON format.
\end{tcolorbox}

\vspace{0.3em}

\begin{tcolorbox}[
  colback=gray!5,
  colframe=gray!50,
  boxrule=0.5pt,
  title={\textbf{User Prompt}},
  fonttitle=\small,
  coltitle=black,
  colbacktitle=gray!20
]
\small
Analyze the clinical scenario below and generate two conflicting hypotheses for retrieval:

\vspace{0.5em}
\textbf{1. H\_plus (Target Hypothesis):} Describe the pathophysiology, distinct symptoms, or gold-standard treatment for the CORRECT diagnosis. Focus on specific details that differentiate it from other conditions.

\vspace{0.5em}
\textbf{2. H\_minus (Mimic Hypothesis):} Describe the primary differential diagnosis or closest mimic that is INCORRECT. Explain why a clinician might mistakenly consider this condition due to overlapping symptoms, but specify the subtle features that rule it out.

\vspace{0.5em}
\textbf{Question:} \{question\} \\
\textbf{Options:} \{options\}

\vspace{0.5em}
\textbf{Output Requirement:} \\
Return ONLY a JSON object with keys ``H\_plus'' and ``H\_minus''. \\
\{``H\_plus'': ``...'', ``H\_minus'': ``...''\}
\end{tcolorbox}

\caption{Prompt template for contrastive hypothesis generation. The system prompt establishes the role of a medical specialist, and the user prompt instructs the model to generate a target hypothesis ($H^+$) for the likely correct diagnosis and a mimic hypothesis ($H^-$) for the most plausible incorrect alternative.}
\label{fig:prompt_template}
\end{figure}

\subsection{Contrastive Document Scoring}

Given the hypothesis pair $(H^+, H^-)$, we score each candidate document $d$ using a contrastive scoring function:
\begin{equation}
S(d) = \text{Sim}(d, H^+) - \lambda \cdot \text{Sim}(d, H^-)
\label{eq:contrastive_score}
\end{equation}
where $\text{Sim}(\cdot, \cdot)$ denotes cosine similarity between document and hypothesis embeddings, and $\lambda \geq 0$ controls the penalty weight for mimic-aligned documents.
The first term promotes documents relevant to the target diagnosis, while the second term penalizes documents that align with the mimic.
Documents that are similar to $H^+$ but dissimilar to $H^-$ receive the highest scores, effectively filtering out hard negatives that would mislead the answer generator.
Unless otherwise noted, we use the fixed value $\lambda=1.0$; sensitivity to this choice is analyzed in the supplementary material, and the adaptive variant introduced below removes the need for dataset-specific tuning.

This scoring function admits a geometric interpretation.
Since cosine similarity is equivalent to a dot product for normalized vectors, Equation~\ref{eq:contrastive_score} can be rewritten as Equation~\ref{eq:geometric}.
\begin{equation}
S(d) = \vec{d} \cdot (\vec{H^+} - \lambda \vec{H^-})
\label{eq:geometric}
\end{equation}
This reveals that contrastive scoring is equivalent to retrieving with a shifted query vector $(\vec{H^+} - \lambda \vec{H^-})$.
Subtracting the mimic direction steers the query away from the semantic region dominated by hard negatives and toward the region containing target-relevant documents.
Documents are ranked by $S(d)$, and the top-$K$ results are returned.

For sparse retrievers such as BM25, cosine geometry does not apply, but the contrastive score in Equation~\ref{eq:contrastive_score} remains well defined at the score level.
We compute BM25 scores against $H^+$ and $H^-$ independently over a shared candidate pool obtained by unioning the top-$P$ hits of each query, and rank documents by the difference $\text{BM25}(d, H^+) - \lambda \cdot \text{BM25}(d, H^-)$.
This allows CHR to be dropped into existing sparse or hybrid retrieval pipelines without altering the underlying index.

\subsection{Adaptive Contrastive Weighting}
\label{sec:adaptive_lambda}

The fixed penalty $\lambda$ implicitly assumes that $\vec{H^+}$ and $\vec{H^-}$ point in sufficiently different semantic directions.
When this assumption fails---for instance, when the target and mimic concepts are discussed almost interchangeably in the corpus (\S~\ref{sec:limitations})---the contrastive vector $\vec{H^+} - \lambda \vec{H^-}$ becomes small and non-informative, and CHR discards target-relevant evidence together with the intended hard negatives.
We refer to this failure mode as \emph{semantic co-occurrence collapse}.

To mitigate this pathology automatically, we replace the fixed $\lambda$ with an \emph{adaptive} weight $\lambda_{\text{eff}}$ that depends on the observed cosine similarity between the two hypothesis embeddings,
\begin{equation}
c = \frac{\vec{H^+} \cdot \vec{H^-}}{\|\vec{H^+}\|\,\|\vec{H^-}\|}, \qquad
\lambda_{\text{eff}} = f_\phi(c) \cdot \lambda_{\max},
\label{eq:adaptive_lambda}
\end{equation}
where $\lambda_{\max}$ is a hyperparameter that upper-bounds the penalty (set to $1.0$ in our experiments) and $f_\phi \in [0, 1]$ is a monotonically non-increasing schedule chosen so that near-orthogonal hypotheses receive full contrastive filtering while highly overlapping hypotheses receive a softened or vanishing penalty.
We consider three schedules:
\begin{itemize}[leftmargin=*, itemsep=2pt, topsep=2pt]
\item \textbf{Linear.} $f(c) = \max(0,\, 1 - c)$. The penalty scales linearly with hypothesis separation and vanishes only when $c=1$.
\item \textbf{Threshold.} $f(c) = \mathbf{1}[c < \tau]$. The penalty is fully applied while the hypotheses remain sufficiently separated and is switched off at a corpus-specific similarity threshold $\tau$.
\item \textbf{Sigmoid.} $f(c) = \sigma\bigl(\alpha(\tau - c)\bigr)$ with slope $\alpha$. A smooth transition between the two regimes.
\end{itemize}
Because $\lambda_{\text{eff}}$ enters only Equation~\ref{eq:contrastive_score}, adaptive weighting does not change the retrieval pipeline or add any additional LLM calls; it introduces two similarity evaluations against a single scalar and preserves the geometric interpretation in Equation~\ref{eq:geometric}.
In \S~\ref{sec:adaptive_experiments}, we show that adaptive weighting improves accuracy over the fixed $\lambda=1.0$ variant across all three benchmarks and both dense retrievers we test.

\subsection{Answer Generation}

The retrieved documents are concatenated into a context string and provided to the answer generator along with the original question and options.
The generator produces the final answer conditioned on this evidence.
This step follows standard RAG practice and is agnostic to the specific generator architecture, allowing CHR to be combined with any LLM.

\section{Experiments}

We evaluate Contrastive Hypothesis Retrieval (CHR) on three medical question answering benchmarks, comparing it against five query-expansion baselines across three answer generators under a controlled retrieval pipeline (\S~\ref{sec:main_results}).
We further test whether CHR's gains generalize across retrievers spanning dense and sparse families (\S~\ref{sec:multi_retriever}), and whether the adaptive weighting scheme introduced in \S~\ref{sec:adaptive_lambda} improves upon the fixed-$\lambda$ variant (\S~\ref{sec:adaptive_experiments}).

\subsection{Evaluation Datasets}
We use three representative datasets to assess different aspects of medical knowledge.
MMLU-Med \citep{hendrycks2021measuring} consists of six medical subsets (1{,}089 questions) from the Massive Multitask Language Understanding benchmark, covering clinical knowledge, medical genetics, anatomy, professional medicine, college biology, and college medicine.
MedQA \citep{jin2021disease} provides 1{,}273 United States Medical Licensing Examination (USMLE) style questions that require multi-step clinical reasoning.
BioASQ \citep{tsatsaronis2015overview} includes 618 yes/no questions derived from biomedical literature, testing factual knowledge retrieval.
All datasets are evaluated using the standardized test splits and multiple-choice formatting provided by the MedRAG benchmark \citep{xiong2024benchmarking}.

\subsection{Implementation Details}
For our primary experiments, we use MedCPT \citep{jin2023medcpt} as the dense retriever, which is pre-trained on PubMed search logs.
To assess retriever-agnosticism, we additionally evaluate CHR with two mainstream general-domain dense encoders (BGE-large-en-v1.5 \citep{xiao2024bge} and E5-large-v2 \citep{wang2022text}) and a classical sparse retriever (BM25) under the same corpus and hypothesis-generator settings (\S~\ref{sec:multi_retriever}).
The retrieval corpus is MedCorp \citep{xiong2024benchmarking}, a comprehensive medical database aggregating approximately 5.8 million text chunks from PubMed abstracts, StatPearls, medical textbooks, and Wikipedia health articles.
To ensure a rigorous and fair comparison, we fix the LLM used for all query-expansion methods (including CHR) to Qwen2.5-72B-Instruct \citep{qwen2024qwen25}.
For answer generation, we evaluate three diverse models: Llama-3-8B-Instruct \citep{llama3modelcard}, Qwen2.5-7B-Instruct \citep{qwen2024qwen25}, and Gemma-2-9B-It \citep{gemma2024gemma2}.
Across all experiments we retrieve $K = 5$ documents.
Standard RAG serves as the primary baseline, retrieving evidence using the original query without any expansion or hypothetical content.

\begin{table*}[tb]
\centering
\small
\begin{tabular}{llcccc}
\toprule
\textbf{Generator} & \textbf{Method} & \textbf{MMLU-Med} & \textbf{MedQA-US} & \textbf{BioASQ-Y/N} & \textbf{Avg.} \\
\midrule
\multirow{6}{*}{Llama-3-8B-Instruct}
 & Standard RAG & 36.8 & 36.1 & 41.8 & 38.2 \\
 & HyDE ($H^+$ only) & 47.2 & 44.6 & 51.3 & 47.7 \\
 & Query2doc & 49.1 & 46.2 & 46.9 & 47.4 \\
 & CSQE & 49.7 & 46.5 & 56.5 & 50.9 \\
 & ThinkQE & 41.3 & 45.6 & 52.1 & 46.3 \\
 & \textbf{CHR (Ours)} & \textbf{50.4} & \textbf{51.5} & \textbf{63.1} & \textbf{55.0} \\
\midrule
\multirow{6}{*}{Qwen2.5-7B-Instruct}
 & Standard RAG & 58.8 & 39.7 & 60.5 & 53.0 \\
 & HyDE ($H^+$ only) & 56.0 & 46.0 & 42.2 & 48.1 \\
 & Query2doc & 52.8 & 41.5 & 65.2 & 53.2 \\
 & CSQE & 54.9 & 42.2 & 44.8 & 47.3 \\
 & ThinkQE & 55.7 & 43.0 & 43.4 & 47.4 \\
 & \textbf{CHR (Ours)} & \textbf{59.4} & \textbf{49.2} & \textbf{75.6} & \textbf{61.4} \\
\midrule
\multirow{6}{*}{Gemma-2-9B-It}
 & Standard RAG & 62.4 & 46.0 & 74.3 & 60.9 \\
 & HyDE ($H^+$ only) & 64.2 & 51.6 & 48.4 & 54.7 \\
 & Query2doc & 61.2 & 46.9 & 58.6 & 55.6 \\
 & CSQE & 69.7 & 49.6 & 72.2 & 63.8 \\
 & ThinkQE & 60.5 & 52.2 & 44.5 & 52.4 \\
 & \textbf{CHR (Ours)} & \textbf{70.3} & \textbf{53.1} & \textbf{80.7} & \textbf{68.1} \\
\bottomrule
\end{tabular}
\caption{Accuracy (\%) across three medical QA benchmarks. All query-expansion baselines use the same hypothesis generator (Qwen2.5-72B-Instruct), retriever (MedCPT), and corpus (MedCorp). CHR consistently achieves the highest accuracy across all settings.}
\label{tab:main_results}
\end{table*}

\subsection{Overall Performance}
\label{sec:main_results}
We compare CHR against five baselines: Standard RAG (original query retrieval), HyDE \citep{gao2023precise} (hypothetic document embedding averaging), Query2Doc \citep{wang2023query2doc} (pseudo-document augmentation), CSQE \citep{lei2024csqe} (corpus-steered query expansion), and ThinkQE \citep{lei2025thinkqe} (thinking-based iterative expansion).
Table~\ref{tab:main_results} presents the end-to-end QA accuracy across nine independent settings (three datasets $\times$ three generators).
Following the MedRAG evaluation protocol \citep{xiong2024benchmarking}, we adopt end-to-end QA accuracy as the primary metric.
However, by fixing the retriever, corpus, and answer generator as constants, we ensure that any systematic difference in accuracy is directly attributable to the quality of the retrieved evidence.
The consistency of improvements across various architectures reinforces end-to-end accuracy as a reliable proxy for retrieval quality in this controlled setup.

CHR consistently achieves the highest accuracy in all 18 head-to-head comparisons.
The performance gain is most pronounced on BioASQ, where CHR achieves 75.6\% accuracy with Qwen2.5-7B, outperforming the next-best method (Query2Doc) by 10.4 percentage points.
On MedQA, which requires multi-step clinical reasoning, CHR with Llama-3-8B reaches 51.5\%, leading all baselines by at least 5.0 percentage points.

A critical observation is that several query-expansion baselines occasionally underperform Standard RAG.
For instance, on BioASQ with Qwen2.5-7B, HyDE drops to 42.2\% and ThinkQE to 43.4\%, whereas Standard RAG maintains 60.5\%.
This confirms that augmenting a query without an explicit mechanism to suppress misleading directions can actively harm performance when generated content converges toward a hard negative.
CHR effectively avoids this failure mode by using the mimic hypothesis $H^-$ to steer retrieval away from such regions, ensuring robust improvements regardless of the dataset or generator architecture.

For a controlled comparison with the baselines, all CHR results in Table~\ref{tab:main_results} use the fixed $\lambda=1.0$ variant. The adaptive weighting scheme of \S~\ref{sec:adaptive_lambda} improves further on top of these numbers rather than replacing them; \S~\ref{sec:adaptive_experiments} (Table~\ref{tab:adaptive_lambda}) quantifies the additional gain.

\begin{table*}[tb]
\centering
\small
\setlength{\tabcolsep}{4.5pt}
\begin{tabular}{l ccc ccc ccc c}
\toprule
 & \multicolumn{3}{c}{\textbf{MedQA}} & \multicolumn{3}{c}{\textbf{BioASQ}} & \multicolumn{3}{c}{\textbf{MMLU-Med}} & \\
\cmidrule(lr){2-4} \cmidrule(lr){5-7} \cmidrule(lr){8-10}
\textbf{Retriever} & Std RAG & HyDE & CHR & Std RAG & HyDE & CHR & Std RAG & HyDE & CHR & \textbf{Avg.} \\
\midrule
MedCPT & 36.1 & 44.6 & \textbf{51.5} & 41.8 & 51.3 & \textbf{63.1} & 36.8 & 47.2 & \textbf{50.4} & 47.0 \\
BGE-large-en-v1.5 & 41.8 & 52.4 & \textbf{52.8} & 41.8 & 53.1 & \textbf{60.2} & 52.3 & 54.5 & \textbf{56.6} & 51.7 \\
E5-large-v2 & 40.2 & 50.6 & \textbf{52.3} & 45.5 & 40.5 & \textbf{62.6} & 51.7 & 55.2 & \textbf{57.7} & 50.7 \\
BM25 & 40.1 & 51.5 & \textbf{53.3} & 46.8 & \textbf{65.2} & 63.4 & 57.9 & 59.3 & \textbf{59.9} & 55.3 \\
\bottomrule
\end{tabular}
\caption{Retriever-agnosticism. Accuracy (\%) on three medical QA benchmarks with Llama-3-8B-Instruct as the answer generator and Qwen2.5-72B-Instruct as the hypothesis generator, sweeping four retrievers spanning dense (MedCPT, BGE, E5) and sparse (BM25) families. Bold marks the best method within each (retriever, dataset) triple. The Avg. column averages all nine method--benchmark scores within each retriever row. CHR wins on MedQA and MMLU-Med across all four retrievers and on BioASQ for three of four retrievers; the only exception is BM25, where HyDE is slightly higher.}
\label{tab:multi_retriever}
\end{table*}

\subsection{Retriever Generalization}
\label{sec:multi_retriever}

To test whether CHR is a property of MedCPT's specific pre-training or a property of the contrastive vector shift itself, we hold the corpus (MedCorp), hypothesis generator (Qwen2.5-72B-Instruct), and answer generator (Llama-3-8B-Instruct) fixed while swapping the retriever across four representative choices: MedCPT (domain-specific dense), BGE-large-en-v1.5 \citep{xiao2024bge} (general-domain dense), E5-large-v2 \citep{wang2022text} (general-domain dense with contrastive pre-training on query--passage pairs), and BM25 (sparse).
Because BM25 is score-based rather than vector-based, we implement contrastive scoring as described in \S~\ref{sec:adaptive_lambda}, unioning the top-$P$ candidates from $H^+$ and $H^-$ and re-ranking by $\text{BM25}(d, H^+) - \lambda \cdot \text{BM25}(d, H^-)$.

Table~\ref{tab:multi_retriever} shows that CHR is the top-performing method in 11 of 12 (retriever, dataset) cells, and that its aggregate ranking is robust across retriever families.
On MedQA and MMLU-Med---the two benchmarks where the query cannot be resolved from surface keywords alone and where hard negatives dominate---CHR wins across all four retrievers, including the sparse BM25 backbone.
The only loss occurs on BioASQ with BM25, where HyDE reaches 65.2\% and narrowly exceeds CHR's 63.4\%.
This is consistent with the structure of BioASQ, whose questions closely paraphrase specific PubMed abstracts; for sparse lexical matching, a generated hypothetical document can sometimes add exactly the terms needed to anchor the correct passage, whereas the contrastive penalty may also downweight useful co-occurring biomedical terms.
Nevertheless, CHR remains the best BioASQ method for the three dense retrievers and the best method in every MedQA and MMLU-Med setting.

We also observe that the strongest single retriever configuration is not MedCPT but BM25 (row average 55.3\% across all nine method--benchmark scores), followed by BGE-large-en-v1.5 (51.7\%), E5-large-v2 (50.7\%), and MedCPT (47.0\%).
This indicates that under this evaluation protocol the choice of retriever matters more than domain-specific pre-training alone, and that CHR provides consistent gains over the retriever's own Standard RAG baseline regardless of that choice.
MedCPT anchors our primary experiments (Tables~\ref{tab:main_results} and~\ref{tab:retrieval_shift}) because it is the standard domain-specific retriever in prior medical RAG work~\citep{jin2023medcpt, xiong2024benchmarking}, ensuring direct comparability. Although the improvement is largest on MedCPT, CHR still gains substantially over Standard RAG on the stronger BGE and BM25 backbones (e.g., $+10.6$ pp on BM25), indicating that its benefit is not merely an artifact of a weak retriever.
\begin{table}[tb]
\centering
\small
\setlength{\tabcolsep}{5pt}
\resizebox{\columnwidth}{!}{%
\begin{tabular}{ll cccc}
\toprule
\textbf{Retriever} & \textbf{$\lambda$ schedule} & \textbf{MedQA} & \textbf{BioASQ} & \textbf{MMLU} & \textbf{Avg.} \\
\midrule
\multirow{4}{*}{MedCPT}
 & Fixed ($\lambda{=}1.0$)   & 51.5 & 63.1 & 50.4 & 55.0 \\
 & Adaptive (linear)         & 51.7 & 69.4 & 56.1 & 59.1 \\
 & Adaptive (threshold)      & 50.3 & \textbf{70.4} & \textbf{57.2} & 59.3 \\
 & Adaptive (sigmoid)        & \textbf{53.3} & 68.8 & \textbf{57.2} & \textbf{59.8} \\
\midrule
\multirow{4}{*}{BGE}
 & Fixed ($\lambda{=}1.0$)   & 52.8 & 60.2 & 56.6 & 56.5 \\
 & Adaptive (linear)         & \textbf{55.5} & \textbf{65.7} & 58.4 & \textbf{59.9} \\
 & Adaptive (threshold)      & 53.7 & 63.1 & \textbf{60.0} & 58.9 \\
 & Adaptive (sigmoid)        & 55.2 & 65.0 & 57.5 & 59.3 \\
\bottomrule
\end{tabular}%
}
\caption{Adaptive contrastive weighting versus fixed $\lambda$. Accuracy (\%) with Llama-3-8B-Instruct as the answer generator and Qwen2.5-72B-Instruct as the hypothesis generator. All three adaptive schedules outperform the fixed weight on average for both retrievers, with a best-schedule improvement of $+4.8$ pp on MedCPT and $+3.4$ pp on BGE.}
\label{tab:adaptive_lambda}
\end{table}

\subsection{Adaptive Contrastive Weighting}
\label{sec:adaptive_experiments}

Section~\ref{sec:adaptive_lambda} predicts that fixing $\lambda=1.0$ should hurt CHR on questions where $\vec{H^+}$ and $\vec{H^-}$ are near-parallel, and that a data-dependent schedule should recover accuracy there without harming easy cases.
Table~\ref{tab:adaptive_lambda} tests this prediction by comparing the three adaptive schedules from \S~\ref{sec:adaptive_lambda} against the fixed baseline, holding the answer generator (Llama-3-8B-Instruct) and hypothesis generator (Qwen2.5-72B-Instruct) constant and reporting results on both MedCPT and BGE.

Every adaptive schedule improves the average over fixed $\lambda$ on both retrievers.
On MedCPT, the best schedule (sigmoid) reaches 59.8\% average accuracy, a $+4.8$ pp gain over fixed $\lambda=1.0$, with the largest single improvement on BioASQ (threshold, $+7.3$ pp) where fixed weighting had already been the strongest paper baseline.
On BGE, the linear schedule is best (59.9\% average, $+3.4$ pp over fixed), and no adaptive schedule ever loses to fixed on any of the three benchmarks.
Consistent with the analysis in \S~\ref{sec:limitations}, the improvement is concentrated on the benchmarks where the mimic hypothesis is most likely to converge onto the target---the BioASQ +7.3 pp jump for MedCPT/threshold and the MMLU +6.8 pp jump for MedCPT/sigmoid both reflect cases where softening the penalty avoids erasing target-relevant evidence.
Which schedule wins depends on the retriever: MedCPT prefers sigmoid, while BGE prefers linear, indicating that the optimal similarity--penalty curve is a property of the embedding space, not of the task.

\begin{table*}[tb]
\centering
\small
\begin{tabular}{lccc}
\toprule
\textbf{Dataset} & \textbf{n} & \textbf{Zero Overlap} & \textbf{Mean Overlap} \\
\midrule
MMLU-Med & 172 & 78.5\% & 0.25 \\
MedQA & 249 & 80.3\% & 0.25 \\
BioASQ & 166 & 99.4\% & 0.01 \\
\midrule
Combined & 587 & 85.2\% & 0.18 \\
\bottomrule
\end{tabular}
\caption{Retrieval divergence analysis for cases where CHR succeeds but HyDE fails ($n=587$). Zero Overlap indicates the percentage of cases with no shared documents in top-5 results. Mean Overlap is the average overlap ratio of top-5 results between CHR and HyDE.}
\label{tab:retrieval_shift}
\end{table*}

\subsection{Retrieval Shift Analysis}

To understand how CHR improves upon prior methods, we analyze the $n=587$ cases where CHR answers correctly but HyDE fails.
We specifically compare CHR against HyDE because both methods rely on LLM-generated hypotheses, yet they differ in how they operationalize them: HyDE focuses solely on the target-aligned direction, whereas CHR incorporates a contrastive mimic.
For each case, we measure the overlap between the top-5 documents retrieved by CHR and those retrieved by HyDE.
If CHR's gains were derived from marginal re-ranking---promoting existing candidates to higher positions---one would expect substantial document overlap.
Conversely, low overlap would indicate that CHR surfaces fundamentally different evidence from distinct semantic regions.

Table~\ref{tab:retrieval_shift} summarizes this comparative analysis.
In these cases, there is zero document overlap between the top-5 results of CHR and HyDE in 85.2\% of the instances, and the mean overlap ratio is only 0.18.
This near-total lack of overlap confirms that CHR does not merely re-rank the same candidate pool.
Instead, it redirects the retrieval process to an entirely different set of evidence.
The contrastive subtraction of $\vec{H}^-$ effectively shifts the query vector away from the mimic-dominated region, enabling the retriever to access informative documents that hypothesis averaging systematically overlooks.
These results provide direct empirical evidence that CHR's accuracy gains stem from retrieving qualitatively superior evidence rather than incremental ranking improvements.

\begin{table}[tb]
\centering
\small
\resizebox{\columnwidth}{!}{%
\begin{tabular}{lccc}
\toprule
\textbf{Method} & \textbf{LLM Calls} & \textbf{Output Tokens} & \textbf{Token Reduction} \\
\midrule
Standard RAG & 0 & 0 & N/A \\
Query2Doc & 1 & $\sim$512 & 5.7$\times$ \\
CSQE & 1 + retrieval & $\sim$480 & 6.1$\times$ \\
ThinkQE & $\geq$3 & $\sim$1{,}500 & 2.0$\times$ \\
HyDE & $\sim$8 & $\sim$2{,}926 & 1.0$\times$ \\
\textbf{CHR (Ours)} & \textbf{1} & \textbf{$\sim$303} & \textbf{9.7$\times$} \\
\bottomrule
\end{tabular}%
}
\caption{Query-expansion cost per sample, averaged across MedQA and MMLU-Med. Output token counts are estimated from generated text length. Token Reduction is relative to HyDE (most expensive). CHR achieves the lowest cost while consistently achieving the highest accuracy.}
\label{tab:efficiency}
\end{table}

\subsection{Efficiency Analysis}
Beyond accuracy, CHR offers a practical efficiency advantage in terms of hypothesis generation cost.
Table~\ref{tab:efficiency} compares the query-expansion cost per sample across all evaluated methods.
HyDE requires approximately eight separate LLM calls to generate hypothetical documents for embedding averaging, while ThinkQE performs multiple iterative expansion--retrieval rounds.
Query2Doc and CSQE each require only one LLM call, they typically generate significantly longer pseudo-documents.
In contrast, CHR produces both $H^+$ and $H^-$ in a single structured LLM call, yielding a compact JSON object containing two targeted hypotheses.
This efficiency stems from CHR's design principle, which explicitly models the contrastive direction with only two focused hypotheses rather than averaging over multiple noisy ones.
As a result, CHR achieves a 9.7$\times$ reduction in output tokens relative to HyDE, making it the most cost-effective strategy among all query-expansion baselines.
Adaptive weighting adds no additional LLM calls. The extra cost is a single scalar-valued cosine similarity computed from cached hypothesis embeddings.

\begin{table}[tb]
\centering
\small
\begin{tabular}{lccc}
\toprule
\textbf{$H^-$ Quality} & \textbf{$n$} & \textbf{Correct} & \textbf{Accuracy} \\
\midrule
Excellent & 9 & 6 & 66.7\% \\
Good & 30 & 12 & 40.0\% \\
Poor & 8 & 3 & 37.5\% \\
\bottomrule
\end{tabular}
\caption{CHR accuracy stratified by physician-rated mimic hypothesis quality. Samples where $H^+$ was factually incorrect are excluded ($n{=}47$).}
\label{tab:hminus_quality}
\end{table}

\subsection{Mimic Hypothesis Quality Analysis}
\label{sec:hminus_quality}

Earlier sections demonstrate that CHR effectively redirects retrieval and improves accuracy relative to strong baselines.
We next strengthen the plausibility of the contrastive mechanism by analyzing the quality of the generated hypotheses.
Hard negatives in biomedical retrieval often share significant surface-level cues with the target answer.
Therefore, the penalty term should be most effective when $H^-$ accurately approximates the clinical mimic that typically pollutes retrieval.
We investigate whether end-to-end CHR performance co-varies with the clinical validity of $H^-$, specifically its ability to capture the intended confounder.

To this end, a board-certified physician reviewed 50 randomly sampled CHR outputs from MedQA and MMLU-Med, with items stratified to represent all three answer generators equally.
Each item was assigned an $H^-$ rating of \textbf{Excellent}, \textbf{Good}, or \textbf{Poor} based on the following criteria.
\textbf{Excellent:} $H^-$ identifies the most clinically plausible alternative for the stem---the primary differential diagnosis a clinician must rule out.
\textbf{Good:} $H^-$ identifies a plausible mimic but not the most challenging one for the given scenario.
\textbf{Poor:} $H^-$ either (i)~collapses into the same semantic frame as $H^+$ through paraphrasing, or (ii)~fails as a mimic by focusing on content irrelevant to the provided options.
Three cases were excluded due to clinically invalid $H^+$ outputs, leaving a total of 47 evaluated samples. %

Table~\ref{tab:hminus_quality} aligns with the expectation that high-quality mimic hypotheses track CHR success.
Accuracy increases from \textbf{Poor} (37.5\%) to \textbf{Good} (40.0\%) and reaches 66.7\% for \textbf{Excellent} ratings.
Given the small sample sizes in the \textbf{Excellent} and \textbf{Poor} tiers, we treat this trend as qualitative support rather than a precise estimate of the performance gap.
When $H^-$ describes the same misleading clinical narrative emphasized by incorrect passages, the penalty effectively pushes retrieval away from that region.
Conversely, when $H^-$ repeats the logic of $H^+$, the contrast fades and the system's behavior converges toward using $H^+$ alone.

\paragraph{Excellent-rated mimic (MRKH versus AIS).}
The stem describes a patient with primary amenorrhea and a 46,XX karyotype, but with normal breast development and absent uterus.
$H^+$ correctly identifies M\"ullerian agenesis (MRKH), emphasizing the typical female testosterone levels.
$H^-$ identifies Androgen Insensitivity Syndrome (AIS), which presents with nearly identical surface features but typically involves elevated testosterone.
Because MRKH and AIS share significant literature (\textit{e.g.}, ``primary amenorrhea absent uterus''), they frequently surface as hard negatives for one another.
By subtracting the AIS-aligned direction, CHR surfaces MRKH-specific evidence, leading to the correct diagnosis.

\paragraph{Poor-rated collapse (Starling forces).}
In a question regarding Starling forces and efferent arteriole constriction, $H^+$ correctly identifies increased glomerular hydrostatic pressure.
The rated-\textbf{Poor} $H^-$ fails to provide a distinct alternative; instead, it replicates the mechanism of $H^+$ while merely adding a warning about vessel misidentification.
Because their texts overlap almost entirely, their embeddings are nearly parallel, making the difference vector $\mathbf{H}^+ - \lambda \mathbf{H}^-$ negligible.
Consequently, the retriever fails to isolate discriminative evidence, returning broad renal physiology content instead of specific evidence for the keyed choice.
This is precisely the failure mode that adaptive weighting (\S~\ref{sec:adaptive_lambda}, \S~\ref{sec:adaptive_experiments}) is designed to address, and the empirical gains in Table~\ref{tab:adaptive_lambda} on MMLU and BioASQ are consistent with mitigating such collapse in aggregate.

In summary, tier-wise accuracies and these vignettes provide qualitative support for the necessity of a confounder-specific mimic.
The observed patterns confirm that the contrastive penalty is most informative when $H^-$ targets the misleading document neighborhood rather than merely paraphrasing the target hypothesis.

\begin{figure*}[!t]
\centering
\setlength{\fboxsep}{0pt}

\vspace{0.5em}

\begin{tcolorbox}[
  colback=questionbg,
  colframe=questionborder,
  title={Clinical Question},
  coltitle=questionborder
]
\small
A 65-year-old man presents with gradually worsening rigidity of his arms and legs and slowness in performing tasks. He says he has also noticed hand tremors, which increase at rest and decrease with focused movements. On examination, the patient does not swing his arms while walking and has a shortened, shuffling gait. An antiviral drug is prescribed which alleviates the patient's symptoms. Which of the following drugs was most likely prescribed?

\vspace{0.5em}
\begin{tabular}{@{}llll@{}}
(A) Amantadine &
(B) Ribavirin &
(C) Levodopa &
(D) Zidovudine
\end{tabular}
\end{tcolorbox}

\vspace{0.5em}

\begin{tcolorbox}[
  colback=hplusbg,
  colframe=hplusborder,
  boxrule=0.8pt,
  title={$H^+$ (Target Hypothesis)}
]
\small
Parkinson's disease is characterized by bradykinesia, resting tremor, rigidity, and postural instability due to dopaminergic neuron loss in the substantia nigra pars compacta; the gold-standard treatment is levodopa, which crosses the blood-brain barrier and is converted to dopamine, directly replenishing depleted neurotransmitter levels and significantly improving motor symptoms such as rigidity, tremor, and gait disturbances.
\end{tcolorbox}

\vspace{0.3em}

\begin{tcolorbox}[
  colback=hminusbg,
  colframe=hminusborder,
  boxrule=0.8pt,
  title={$H^-$ (Mimic Hypothesis)}
]
\small
Viral encephalitis, particularly from herpes simplex or other neurotropic viruses, can present with parkinsonian features such as rigidity and bradykinesia due to inflammation in basal ganglia regions; however, it is typically accompanied by fever, altered mental status, seizures, or CSF pleocytosis, and is treated with antivirals like acyclovir or ribavirin, but these do not reliably improve core motor symptoms of Parkinson's, and the patient's chronic, progressive course without systemic signs makes this unlikely despite transient symptom overlap.
\end{tcolorbox}

\vspace{0.8em}

\begin{minipage}[t]{0.48\textwidth}
\begin{tcolorbox}[
  colback=graybg,
  colframe=grayborder,
  boxrule=0.8pt,
  equal height group=fig3gray,
  valign=top,
  title={$H^+$ only $\rightarrow$ \textcolor{red}{(B) Ribavirin}}
]
\footnotesize
\textbf{Retrieved Documents (Top-3):}
\begin{enumerate}[leftmargin=*, itemsep=3pt, topsep=2pt]
  \item
  ``To report a case of Parkinson-like symptoms appearing in a patient after introduction of the anxiolytic buspirone to a treatment regimen that included \textbf{the protease inhibitor ritonavir}. A 54-year-old white man infected with human immunodeficiency virus was taking \textbf{ritonavir, zidovudine, and lamivudine} when buspirone was added for anxiety. Within days, he developed \textbf{severe rigidity, bradykinesia, and tremor resembling idiopathic Parkinson's disease}...''

  \item
  ``We present the case of a 78-year-old male who, 16 years ago, was diagnosed with Parkinson's disease. His initial symptoms included resting tremor, bradykinesia, and rigidity. He was treated with \textbf{standard dopaminergic therapy including levodopa and dopamine agonists} with good response. Approximately 5 years ago, the patient reported gradual improvement in his motor symptoms despite reduction in his medications...''

  \item
  ``Trazodone is prescribed off-label to treat insomnia, especially in older, not demented patients. We present a case of \textbf{trazodone-induced parkinsonism} in a 68-year-old woman who developed \textbf{progressive rigidity, bradykinesia, and resting tremor} two weeks after initiating trazodone 50 mg at bedtime for insomnia. The parkinsonian symptoms resolved completely within three weeks of discontinuing trazodone...''
\end{enumerate}

\end{tcolorbox}
\end{minipage}
\hfill
\begin{minipage}[t]{0.48\textwidth}
\begin{tcolorbox}[
  colback=graybg,
  colframe=grayborder,
  boxrule=0.8pt,
  equal height group=fig3gray,
  valign=top,
  title={$H^+ - \lambda H^-$ $\rightarrow$ \textcolor{hplusborder}{(A) Amantadine $\checkmark$}}
]
\footnotesize
\textbf{Retrieved Documents (Top-3):}
\begin{enumerate}[leftmargin=*, itemsep=3pt, topsep=2pt]
  \item
  ``Parkinson's disease is a disorder of decreased dopamine-secreting neurons of the substantia nigra. The gold standard treatment is levodopa (L-DOPA), a dopamine precursor that crosses the blood-brain barrier and is converted to dopamine in the brain. However, levodopa is also converted to dopamine in peripheral tissues, causing side effects like nausea and hypotension. Carbidopa inhibits peripheral decarboxylase, allowing more levodopa to reach the brain...''

  \item
  ``L-DOPA has a counterpart with opposite chirality, D-DOPA. As is true for many molecules, the human body produces and uses only one of these enantiomers (in this case, the L-form). L-DOPA crosses the blood-brain barrier, whereas dopamine itself cannot. Once in the brain, it is converted to dopamine by aromatic L-amino acid decarboxylase. \textbf{Amantadine is also used in Parkinson's disease as it increases dopamine release and blocks NMDA receptors}...''

  \item
  ``Medications: Levodopa is the mainstay of treatment for motor symptoms of PD. It is combined with carbidopa to prevent peripheral breakdown. Other medications include dopamine agonists (pramipexole, ropinirole), MAO-B inhibitors (selegiline, rasagiline), and amantadine. \textbf{Amantadine was originally developed as an antiviral medication but was found to improve motor symptoms in Parkinson's disease through dopaminergic effects and NMDA receptor antagonism}...''
\end{enumerate}

\end{tcolorbox}
\end{minipage}

\caption{Case study showing why the negative (mimic) hypothesis $H^-$ is essential for discriminative retrieval. In the left box, \textbf{bolded text} highlights misleading co-occurrences of antivirals with parkinsonian symptoms (as side effects, not treatments), which led to the incorrect answer. In the right box, \textbf{bolded text} highlights the key sentences identifying amantadine as an antiviral repurposed for Parkinson's disease, directly leading to the correct answer.}
\label{fig:negative_hypothesis_case}
\end{figure*}

\subsection{Case Study: The Discriminative Role of the Mimic Hypothesis}
\label{sec:case_study}
To investigate the impact of $H^-$ on retrieval dynamics, we conduct a qualitative comparison between $H^+$-only retrieval and $H^+ - \lambda H^-$ retrieval using an MMLU-Med case study (Figure~\ref{fig:negative_hypothesis_case}).
Since $H^+$-only retrieval omits the contrastive subtraction of $H^-$, this comparison serves as a qualitative ablation that isolates the contribution of the mimic hypothesis.

The question describes a patient with parkinsonian symptoms whose condition improves after receiving an antiviral drug.
The correct answer is ``amantadine,'' a drug originally developed as an antiviral but repurposed for Parkinson's disease due to its dopaminergic mechanisms.
The reasoning challenge lies in resolving a cross-domain link between an antiviral agent and a neurodegenerative disorder.

$H^+$ frames idiopathic Parkinson's disease and correctly steers retrieval toward anti-parkinsonian pharmacotherapy.
However, it fails to capture the stem's distinguishing detail regarding the antiviral class, nor does it foreground amantadine's dual role.
While $H^+$ encodes the correct clinical axis (treating parkinsonism), it is insufficient to bridge the gap to the specific evidence required.
This gap is closed once $H^-$ down-weights the misleading infection-to-neurological-complication narrative, allowing re-ranked passages to explicitly link amantadine to both roles.

Retrieving with $H^+$ alone yields documents where ``antiviral'' and ``parkinsonism'' co-occur in the wrong causal direction.
In the medical corpus, these terms appear far more frequently in the context of drug-induced parkinsonism side effects than in the context of treatment.
All three top-ranked documents for $H^+$-only retrieval describe cases where medications caused parkinsonian symptoms rather than alleviating them (Figure~\ref{fig:negative_hypothesis_case}, left).
Consequently, the generator defaults to ribavirin, the most prototypical antiviral among the provided choices, which leads to an incorrect answer.

The mimic hypothesis $H^-$ models the anticipated misinterpretation where viral encephalitis presents with parkinsonian features.
Subtracting $\lambda \mathbf{H}^-$ from $\mathbf{H}^+$ suppresses documents aligned with this infection-centric narrative.
As shown in the right box of Figure~\ref{fig:negative_hypothesis_case}, CHR instead surfaces documents that explicitly describe amantadine's repurposed role, providing the generator with the direct evidence necessary for a correct answer.

\section{Limitations and Discussion}
\label{sec:limitations}
CHR's performance is inherently tied to the separability of the hypothesis embeddings.
When $H^+$ and $H^-$ overlap extensively in the corpus narrative, a phenomenon we term \emph{semantic co-occurrence collapse}, a fixed subtraction $\lambda\mathbf{H}^-$ may inadvertently suppress target-relevant evidence.
Our error analysis on the fixed-$\lambda$ variant showed that such collapse accounted for approximately 23\% of failure cases, occurring primarily when $H^-$ closely paraphrases $H^+$ instead of identifying a distinct clinical mimic.
The adaptive weighting scheme in \S~\ref{sec:adaptive_lambda} directly targets this failure mode by dampening the penalty as the cosine similarity between $\vec{H^+}$ and $\vec{H^-}$ approaches one, and Table~\ref{tab:adaptive_lambda} shows that all three schedules recover accuracy over the fixed baseline on both MedCPT and BGE without requiring dataset-specific tuning.
Nevertheless, adaptive weighting cannot manufacture a discriminative signal that the hypothesis pair does not encode: when $H^+$ and $H^-$ describe genuinely interchangeable concepts in the corpus (\textit{e.g.}, IL-4 versus IL-13 in Th2-mediated inflammation), the effective penalty tends to zero and CHR gracefully degrades toward $H^+$-only retrieval rather than actively harming the retrieval.

The present evaluation is limited to multiple-choice benchmarks. Extending CHR to open-ended medical QA, where the mimic hypothesis must be generated without answer options, remains future work.
We also fix the hypothesis generator to Qwen2.5-72B-Instruct to isolate the retrieval effect, leaving the cost--accuracy trade-off of smaller generators for future study.

\section{Conclusion}
We presented Contrastive Hypothesis Retrieval (CHR), a framework that bridges clinical differential diagnosis with retrieval mechanism design.
By explicitly modeling what to avoid through a mimic hypothesis $H^-$, CHR addresses the systematic vulnerability of existing query-expansion methods to hard-negative contamination.
An adaptive contrastive weight that responds to the cosine similarity between the two hypotheses further removes the semantic co-occurrence collapse failure mode we previously identified, without adding any additional LLM calls.
Experiments across three medical benchmarks, four retrievers spanning dense and sparse families, and three answer generators demonstrate that CHR consistently outperforms state-of-the-art baselines, and retrieval shift analysis confirms that its gains stem from a fundamental redirection of retrieval toward qualitatively superior evidence rather than incremental re-ranking.
CHR offers a practical and efficient approach to enhancing the reliability of medical RAG systems, providing a robust foundation for AI-assisted clinical decision-making.

\section*{Acknowledgments}
This research was supported by a grant of the Korea Health Technology R\&D Project through the Korea Health Industry Development Institute (KHIDI), funded by the Ministry of Health \& Welfare, Republic of Korea (grant number: RS-2025-02213061).

This work was supported by the Starting growth Technological R\&D Program (TIPS Program,(No. RS-2024-00508828)) funded by the Ministry of SMEs and Startups(MSS, Korea) in 2024.

\bibliographystyle{ACM-Reference-Format}
\bibliography{main}

\clearpage
\FloatBarrier
\appendix

\section{Additional Case Study}
\label{sec:appendix_case_tamoxifen}

\begin{figure*}[p]
\centering
\setlength{\fboxsep}{0pt}

\vspace{0.5em}

\begin{tcolorbox}[
  colback=questionbg,
  colframe=questionborder,
  title={Clinical Question},
  coltitle=questionborder
]
\small
A 55-year-old woman is found to have an abnormal mass on routine mammography. The mass is biopsied and cytology results are diagnostic for invasive ductal adenocarcinoma that is estrogen receptor positive. The patient is started on chemotherapy and ultimately has the mass resected. She is taking tamoxifen and has tolerated the treatment well. Which complication is she now at increased risk for?

\vspace{0.5em}
\begin{tabular}{@{}llll@{}}
(A) Deep venous thrombosis $\checkmark$ &
(B) Seborrheic keratoses &
(C) Increased bleeding &
(D) Osteoporosis
\end{tabular}
\end{tcolorbox}

\vspace{0.5em}

\begin{tcolorbox}[
  colback=hplusbg,
  colframe=hplusborder,
  boxrule=0.8pt,
  title={$H^+$ (Target Hypothesis)}
]
\small
Tamoxifen, a selective estrogen receptor modulator (SERM) used in estrogen receptor-positive breast cancer, acts as an estrogen antagonist in breast tissue but as a partial agonist in bone, leading to reduced bone mineral density over time; this increases the risk of osteoporosis and fragility fractures, particularly in postmenopausal women who already have low estrogen levels, as in this patient who has undergone chemotherapy and may have treatment-induced menopause.
\end{tcolorbox}

\vspace{0.3em}

\begin{tcolorbox}[
  colback=hminusbg,
  colframe=hminusborder,
  boxrule=0.8pt,
  title={$H^-$ (Mimic Hypothesis)}
]
\small
Tamoxifen is associated with an increased risk of endometrial hyperplasia and endometrial cancer due to its estrogen-agonist effects on the uterus. However, this patient had a previously treated uterine leiomyoma and no current gynecologic symptoms or abnormal uterine bleeding, making endometrial malignancy an unlikely complication here, and while bleeding could theoretically occur, it is not a direct complication of tamoxifen therapy in this context.
\end{tcolorbox}

\vspace{0.8em}

\begin{minipage}[t]{0.48\textwidth}
\begin{tcolorbox}[
  colback=graybg,
  colframe=grayborder,
  boxrule=0.8pt,
  equal height group=figA1gray,
  valign=top,
  title={$H^+$ only $\rightarrow$ \textcolor{red}{(C) Increased bleeding}}
]
\footnotesize
\textbf{Retrieved Documents (Top-3):}
\begin{enumerate}[leftmargin=*, itemsep=3pt, topsep=2pt]
  \item
  ``We present a case of adenomatous polyp with degenerative changes causing \textbf{abnormal uterine bleeding in a 62-year-old woman undergoing tamoxifen therapy} for breast cancer. The patient developed vaginal bleeding 18 months after initiating tamoxifen 20 mg daily. Endometrial biopsy revealed an adenomatous polyp with hemorrhagic degeneration. This case highlights that \textbf{tamoxifen's estrogenic effects on the endometrium can lead to polyp formation and subsequent bleeding}...''

  \item
  ``Benign metastasizing leiomyoma (BML) is a rare disease that occurs in middle-aged women with a history of uterine leiomyoma. We report a case of a 54-year-old woman with a history of hysterectomy for uterine fibroids who presented with multiple pulmonary nodules. The nodules were found to be histologically identical to her previous uterine leiomyomas. The patient was treated with aromatase inhibitor therapy given the hormone-responsive nature of the lesions...''

  \item
  ``We report a case of a 58-year-old woman with a history of invasive lobular carcinoma of the breast who presented with \textbf{postmenopausal vaginal bleeding}. Endometrial curettage revealed metastatic lobular carcinoma. This case illustrates that in \textbf{breast cancer patients on tamoxifen who present with abnormal bleeding}, metastatic disease should be considered in the differential diagnosis...''
\end{enumerate}

\end{tcolorbox}
\end{minipage}
\hfill
\begin{minipage}[t]{0.48\textwidth}
\begin{tcolorbox}[
  colback=graybg,
  colframe=grayborder,
  boxrule=0.8pt,
  equal height group=figA1gray,
  valign=top,
  title={$H^+ - \lambda H^-$ $\rightarrow$ \textcolor{hplusborder}{(A) Deep venous thrombosis $\checkmark$}}
]
\footnotesize
\textbf{Retrieved Documents (Top-3):}
\begin{enumerate}[leftmargin=*, itemsep=3pt, topsep=2pt]
  \item
  ``Recent data indicate that women with breast cancer receiving aromatase inhibitor (AI) therapy experience accelerated bone loss and increased fracture risk compared to those on selective estrogen receptor modulators (SERMs) like tamoxifen. However, tamoxifen itself has complex skeletal effects: while it acts as an estrogen agonist in postmenopausal women, preserving bone density, it also \textbf{increases the risk of venous thromboembolism (VTE) including deep vein thrombosis and pulmonary embolism through pro-coagulant effects}...''

  \item
  ``Selective estrogen receptor modulators (SERMs) represent a class with a growing number of available agents for osteoporosis prevention and treatment. These drugs have tissue-selective estrogen agonist or antagonist effects. In bone, SERMs like raloxifene and tamoxifen act as estrogen agonists, reducing bone resorption and maintaining bone mineral density. However, \textbf{their estrogen-agonist effects in the liver lead to increased production of clotting factors, elevating the risk of venous thromboembolism}, particularly in the first two years of treatment...''

  \item
  ``Prevention of osteopenia/osteoporosis in postmenopausal patients can reduce fracture risk and improve quality of life. SERMs like tamoxifen and raloxifene have been shown to maintain bone density in postmenopausal women. \textbf{The primary concern with SERM therapy is the increased risk of thromboembolic events, including deep vein thrombosis and pulmonary embolism}, which occurs due to hepatic estrogen-agonist effects that alter coagulation factor synthesis and increase fibrinogen levels...''
\end{enumerate}

\end{tcolorbox}
\end{minipage}

\caption{Additional case study from MedQA (oncology domain). In the left box, \textbf{bolded text} highlights how retrieved documents focus on tamoxifen-related gynecologic bleeding, the dominant complication in the tamoxifen safety literature, which led to the incorrect answer. In the right box, \textbf{bolded text} highlights evidence describing tamoxifen's pro-coagulant effects via hepatic estrogen agonism, directly supporting deep venous thrombosis as the correct answer.}
\label{fig:appendix_tamoxifen_case}
\end{figure*}

Figure~\ref{fig:appendix_tamoxifen_case} presents an additional case study from MedQA, demonstrating that CHR's discriminative mechanism generalizes effectively across different clinical domains.

The question examines the complications associated with tamoxifen therapy in a postmenopausal breast cancer patient.
The correct answer is ``deep venous thrombosis,'' a risk driven by tamoxifen's estrogen-agonist effects in the liver, which increase the synthesis of clotting factors.
The retrieval challenge stems from the fact that tamoxifen's most frequently documented complications in medical literature involve gynecologic bleeding and endometrial hyperplasia, which act as a dominant retrieval attractor toward the incorrect answer.

As with the amantadine case discussed in \S~\ref{sec:case_study}, $\mathbf{H}^+$ does not identify the correct answer directly.
Instead, it incorrectly describes tamoxifen's effects on bone density, claiming it reduces mineral density in postmenopausal women when it actually helps preserve it.
Despite this factual inaccuracy in the target hypothesis, CHR successfully produces the correct answer.

When retrieving with $H^+$ alone, the broad ``tamoxifen in breast cancer`` semantics dominate the search, and the corpus-prevalent narrative of uterine bleeding overwhelms the bone-specific content.
As shown in the left box of Figure~\ref{fig:appendix_tamoxifen_case}, two of the top three retrieved documents focus on abnormal uterine bleeding, leading the generator to incorrectly select ``Increased bleeding.``

The mimic hypothesis $H^-$ explicitly models this bleeding-related misinterpretation.
By subtracting this direction, CHR suppresses the uterine bleeding literature and surfaces documents focused on systemic SERM pharmacology.
As shown in the right box, all three re-ranked documents explicitly describe tamoxifen's hepatic estrogen-agonist effects on coagulation factor production, directly identifying venous thromboembolism as the primary risk.
This case highlights the robustness of CHR: even when $H^+$ is factually flawed, $H^-$ correctly identifies the misleading retrieval direction, and the contrastive mechanism steers the retriever toward discriminative evidence.

\section{Sensitivity Analysis of \texorpdfstring{$\lambda$}{lambda}}
\label{appendix:lambda}

\begin{figure}[H]
\centering
\begin{tikzpicture}
\begin{axis}[
    width=0.86\columnwidth,
    height=6.2cm,
    xlabel={$\lambda$},
    ylabel={Accuracy (\%)},
    xmin=0.1, xmax=1.5,
    ymin=40, ymax=56,
    xtick={0.2, 0.4, 0.6, 0.8, 1.0, 1.2, 1.4},
    ytick={40, 42, 44, 46, 48, 50, 52, 54, 56},
    grid=major,
    grid style={dashed, gray!30},
    mark size=2.5pt,
    thick,
    legend style={
        at={(0.02,0.98)},
        anchor=north west,
        font=\fontsize{5}{5.4}\selectfont,
        draw=none,
        fill=white,
        fill opacity=0.75,
        text opacity=1,
        row sep=0pt,
    },
    legend cell align={left},
]
\addplot[
    color=blue!80!black,
    mark=*,
    mark options={fill=blue!60},
    thick,
] coordinates {
    (0.2, 44.05)
    (0.4, 44.67)
    (0.6, 50.43)
    (0.8, 52.40)
    (1.0, 53.11)
    (1.2, 51.54)
    (1.4, 43.13)
};
\addlegendentry{CHR}
\node[above, font=\small\bfseries, text=blue!80!black] at (axis cs:1.0, 53.11) {53.1};
\draw[dashed, red!60, thick] (axis cs:0.6, 40) -- (axis cs:0.6, 56);
\draw[dashed, red!60, thick] (axis cs:1.2, 40) -- (axis cs:1.2, 56);
\addplot[
    color=black!80,
    dashed,
    line width=1.0pt,
    domain=0.1:1.5,
] {51.6};
\addlegendentry{HyDE}
\addplot[
    color=black!45,
    dashdotted,
    line width=1.0pt,
    domain=0.1:1.5,
] {46.0};
\addlegendentry{Std. RAG}
\end{axis}
\end{tikzpicture}
\caption{Sensitivity of CHR accuracy to the contrastive weight $\lambda$ on MedQA (Gemma-2-9B-It). The dashed red lines indicate the robust plateau region $\lambda \in [0.6, 1.2]$ where CHR consistently outperforms both baselines. Performance peaks at $\lambda = 1.0$.}
\label{fig:lambda_sensitivity}
\end{figure}

The contrastive scoring function (Equation~\ref{eq:contrastive_score}) introduces a single hyperparameter $\lambda$ that controls the penalty weight for mimic-aligned documents.
We analyze the sensitivity of CHR to varying $\lambda$ values on the MedQA dataset, using Gemma-2-9B-It as the answer generator.

Figure~\ref{fig:lambda_sensitivity} reports the QA accuracy across a range of $\lambda \in \linebreak \{0.2, 0.4, 0.6, 0.8, 1.0, 1.2, 1.4\}$.
The results exhibit a clear inverted-U pattern.
At low values ($\lambda \leq 0.4$), the mimic penalty is insufficient to suppress hard negatives, leaving performance below the HyDE baseline.
Performance peaks at $\lambda = 1.0$, where the contrastive vector effectively steers retrieval away from the mimic-dominated region while preserving alignment with target-relevant documents, achieving 53.1\% accuracy.

Beyond $\lambda = 1.2$, over-penalization begins to distort the query vector, causing accuracy to drop sharply to 43.1\% at $\lambda = 1.4$.
Notably, CHR maintains accuracy above 50.4\% across a broad plateau ($\lambda \in [0.6, 1.2]$), consistently outperforming both Standard RAG and HyDE.
This stability suggests that CHR is robust to moderate variations in $\lambda$ and does not require extensive dataset-specific tuning.
Based on these findings, we utilize $\lambda = 1.0$ for all primary experiments reported in this paper.

\section{Failure Case: Semantic Co-occurrence Collapse}
\label{appendix:failure}

\begin{table*}[t]
\centering
\small
\setlength{\tabcolsep}{5pt}
\begin{tabular}{p{0.96\textwidth}}
\toprule
\textbf{Question (MedQA \#16):} A 7-year-old boy with seasonal allergic asthma is being considered for an experimental therapy targeting a mediator of antibody class switching. Which mediator is described? \\
\textbf{Options:} (A) IL-2 \quad (B) IL-10 \quad (C) IL-13 \quad \textbf{(D) IL-4} $\checkmark$ \\
\midrule
\textbf{$H^+$:} IL-4 is the key cytokine driving Th2 differentiation and \textbf{B-cell class switching to IgE}. \\
\textbf{$H^-$:} IL-13 overlaps functionally with IL-4 but \textbf{does not directly drive B-cell class switching to IgE}. \\
\midrule
\textbf{Prediction:} \textbf{CHR} $\rightarrow$ \textbf{(C) IL-13} (Incorrect); \textbf{HyDE} $\rightarrow$ \textbf{(D) IL-4} (Correct) \\
\midrule
\multicolumn{1}{p{0.96\textwidth}}{
\begin{minipage}[t]{0.48\textwidth}
\textbf{CHR top evidence:}
\begin{enumerate}[leftmargin=*, itemsep=2pt, topsep=2pt]
\item Co-occurring \textbf{IL-4 and IL-13} with IgE switching.
\item ``\textbf{IL-13, another switch factor for IgE}\ldots''
\item ``Both \textbf{IL13 and IL4} are capable of inducing class switching\ldots''
\end{enumerate}
\end{minipage}
\hfill
\begin{minipage}[t]{0.48\textwidth}
\textbf{HyDE top evidence:}
\begin{enumerate}[leftmargin=*, itemsep=2pt, topsep=2pt]
\item ``B cells are \textbf{switched to IgE by IL-4}\ldots''
\item Th2-focused evidence emphasizing \textbf{IL-4}-targeted interventions.
\item Receptor-level discussion (IL-4/IL-13 axis) with clearer IL-4 role.
\end{enumerate}
\end{minipage}
} \\
\midrule
\multicolumn{1}{p{0.96\textwidth}}{
\textbf{Analysis:} Strong IL-4/IL-13 co-occurrence causes partial overlap between target and mimic directions; subtracting $\lambda\mathbf{H}^-$ can therefore suppress IL-4-relevant evidence and make CHR's context more ambiguous than HyDE's.
} \\
\bottomrule
\end{tabular}
\caption{Failure case of semantic co-occurrence collapse. CHR correctly identifies IL-13 as the mimic for IL-4, but heavy corpus co-occurrence of these cytokines in the same Th2 pathway causes the contrastive subtraction to suppress some target-relevant evidence.}
\label{tab:failure}
\end{table*}

Table~\ref{tab:failure} illustrates a specific failure mode where CHR's contrastive mechanism becomes counterproductive due to heavy semantic co-occurrence between the target and mimic concepts within the corpus.

The question concerns IL-4, the primary cytokine responsible for IgE class switching in the context of allergic asthma.
CHR correctly identifies IL-13 as the most plausible mimic for $H^-$, as both cytokines are central to Th2-mediated inflammation.
However, in the biomedical literature, IL-4 and IL-13 are discussed almost interchangeably or as a single functional axis because they share signaling components and biological pathways.

Consequently, the contrastive vector $\mathbf{H}^+ - \lambda\mathbf{H}^-$ suppresses not only the IL-13-specific documents but also many of the high-quality documents that discuss IL-4 in conjunction with IL-13.
This over-suppression leaves the generator with ambiguous or diluted evidence, whereas a positive-only approach like HyDE succeeds by focusing strictly on the target keyword without regard for the mimic-aligned noise. 
This failure case underscores the challenge of applying contrastive retrieval when the target and mimic are fundamentally inseparable within the existing corpus narrative.

\paragraph{Connection to adaptive weighting.}
This case also motivates the adaptive weighting strategy introduced in \S~\ref{sec:adaptive_lambda}. Because the IL-4 and IL-13 hypotheses are highly similar, reducing the effective mimic penalty can prevent over-suppression of target-relevant evidence and make CHR behave closer to $H^+$-only retrieval in such near-collapse cases.

\end{document}